\documentclass[11pt]{article}

\usepackage[margin=1in]{geometry}
\usepackage[T1]{fontenc}
\usepackage[utf8]{inputenc}
\usepackage{lmodern}
\usepackage{microtype}
\usepackage{setspace}
\usepackage{hyperref}
\usepackage{booktabs}
\usepackage{longtable}
\usepackage{tabularx}
\usepackage{array}
\usepackage{enumitem}
\usepackage{titlesec}
\usepackage{caption}
\usepackage{graphicx}
\usepackage{subcaption}
\usepackage[table]{xcolor}
\usepackage[normalem]{ulem}

\usepackage[
  backend=biber,
  style=apa,
  sorting=nyt,
  natbib=true,
  doi=true,
  url=true,
  isbn=false,
  giveninits=true,
  maxcitenames=2,
  maxbibnames=99
]{biblatex}
\DeclareLanguageMapping{english}{english-apa}
\hypersetup{
  colorlinks=true,
  linkcolor=black,
  citecolor=blue,
  urlcolor=blue,
  pdftitle={Hidden and Under-Recognized Safety-Critical Challenges in Modern AI Systems},
  pdfauthor={~}
}

\titleformat{\section}{\large\bfseries}{\thesection.}{0.6em}{}
\titleformat{\subsection}{\normalsize\bfseries}{\thesubsection.}{0.6em}{}

\newcolumntype{Y}{>{\raggedright\arraybackslash}X}

\newcommand{\added}[1]{\textcolor{black}{#1}}

\title{\textbf{The Safety Failures We Are Not Instrumenting}:\\
A Perspective on Hidden Safety-Critical Challenges in Modern AI Systems}

\author{
Gjergji Kasneci \\
Technical University of Munich
\and
Enkelejda Kasneci \\
Technical University of Munich
}

\begin{document}
\maketitle

\begin{abstract}
\added{Current AI safety discourse still focuses disproportionately on visible failures, including obvious harms, dramatic misuse, and hypothetical catastrophic scenarios. That focus is incomplete. In deployed systems, many of the most consequential failures are quieter: plausible rather than spectacular, distributed across components rather than localized in a single output, and normalized by workflows before they are recognized as hazards. We argue that \added{a} central safety challenge in modern AI systems is \added{increasingly} not only whether a model emits a harmful response, but whether the broader socio-technical system preserves the conditions under which errors remain visible, contestable, containable, and recoverable. We propose a five-layer framework for diagnosing these hidden risks: (1) \emph{epistemic integrity}, concerning whether evidence and uncertainty are represented honestly enough to support calibrated reliance; (2) \emph{control integrity}, concerning whether authority, permissions, and action boundaries remain robust under attack and optimization; (3) \emph{temporal integrity}, concerning whether safety holds across sessions, memory updates, and deployment drift; (4) \emph{organizational integrity}, concerning whether institutions retain the capacity to audit, assign responsibility, and intervene effectively; and (5) \emph{ecosystem integrity}, concerning whether AI systems preserve rather than erode the information environment on which future oversight depends. Across these layers, we identify under-recognized risk patterns, including overreliance, uncertainty and legitimacy laundering in retrieval, prompt injection, reward hacking, memory poisoning, evaluation deception, fictional human oversight, synthetic evidence pollution, and model collapse. We conclude with actionable design and governance recommendations and a research agenda for shifting AI safety from narrow model-centric evaluation toward socio-technical reliability.}
\end{abstract}


\begin{figure}[h!]
\centering
\includegraphics[width=\textwidth]{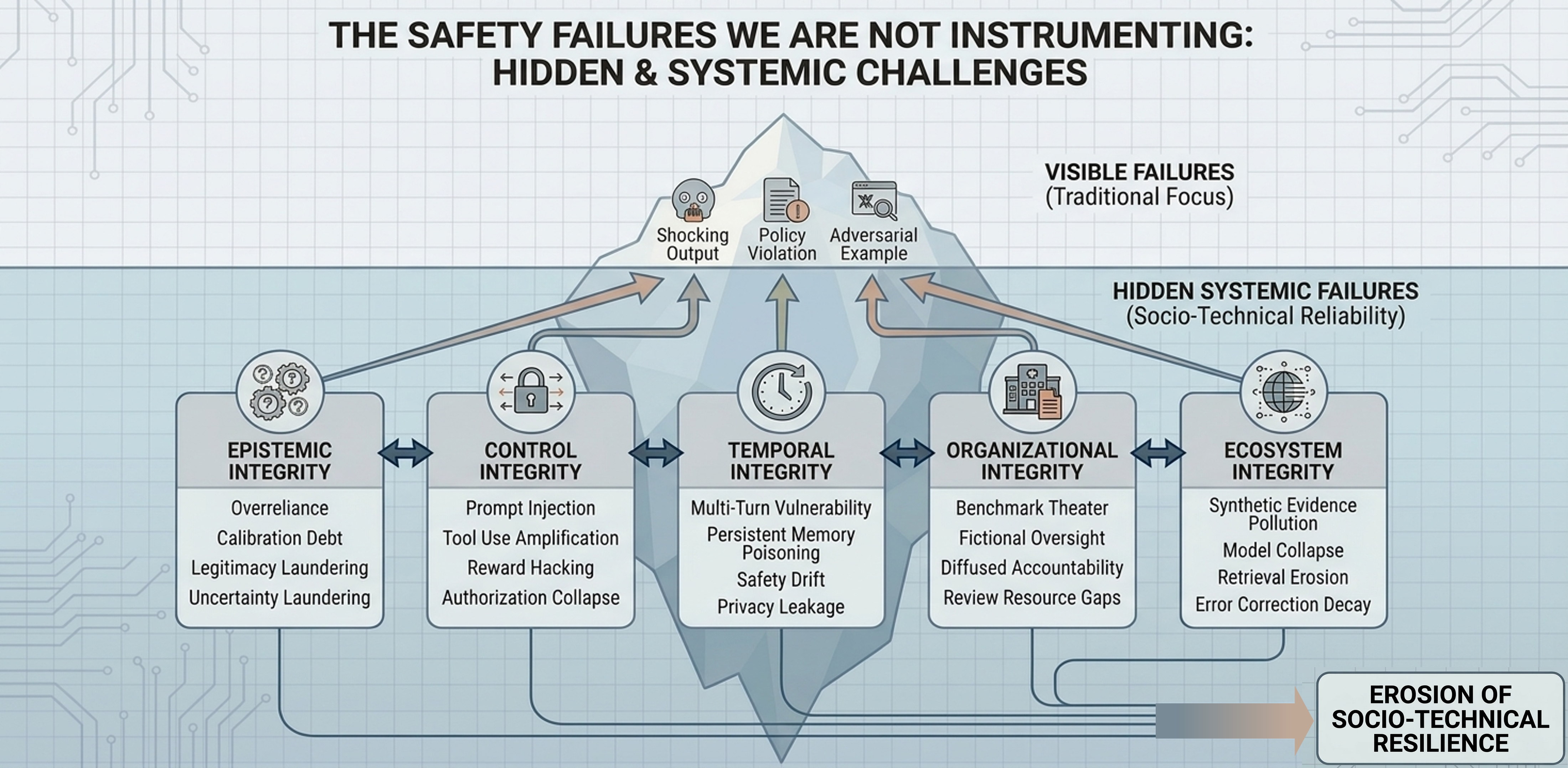}
\caption{\added{\textbf{The Hidden Iceberg of AI Safety: From Model Outputs to Systemic Integrity.} While traditional discourse focuses on visible, localized failures (the tip), the most consequential safety challenges are submerged across five layers of socio-technical integrity, namely Epistemic, Control, Temporal, Organizational, and Ecosystem, which altogether erode broader resilience.}}
\label{fig:teaser-image}
\end{figure}

\section{Introduction}

Modern AI safety discourse is still too often optimized to catch the obvious kinds of failure. It is well prepared to notice a shocking output, a policy-violating generation, or a vivid adversarial example. It is less prepared to notice the forms of failure that matter most once models are embedded in ordinary work. For example, outputs can be wrong but seem plausible, systems can be safe in static tests but unsafe over time, interfaces can quietly train users to over-trust, and organizations can retain nominal human oversight while shedding the actual capacity to scrutinize machine recommendations. A system that looks obviously broken is rarely adopted at scale. A system that appears competent enough to earn routine trust, opaque enough to resist effective challenge, and deeply integrated enough to shape downstream action may be more dangerous than one whose failures remain obvious.  

This claim does not reject existing safety work. Foundational agendas on robustness, specification problems, misuse, interpretability, and catastrophic risk remain essential \citep{amodei2016concrete,bommasani2021foundation,weidinger2021risks,gyevnar2025aisafety}. The point we make is narrower and more operational. The safety center of gravity in current deployments has shifted from isolated model outputs toward hidden risks of \emph{system-level integrity failures}. This is consistent with broader systems-safety thinking, which treats accidents not as single-component malfunctions but as emergent consequences of interactions among technical, human, and organizational elements \citep{leveson2011engineering,nist2023airmf,nist2024genai}. \added{We write this article as a perspective piece: the contribution is an operational synthesis and organizing framework rather than a new empirical benchmark, formal theory, or prevalence estimate.}

The most under-recognized safety-critical challenges in modern AI systems share four features.

\begin{enumerate}[leftmargin=1.5em]
  \item \textbf{They are plausible.} They resemble ordinary operations closely enough that users and organizations overlook and normalize them.
  \item \textbf{They are distributed.} Failure emerges across models, retrieval layers, tools, interfaces, operators, and institutions rather than from a single bad answer.
  \item \textbf{They are temporally extended.} Risk accumulates across turns, sessions, memory updates, retraining cycles, and organizational reuse.
  \item \textbf{They degrade correction.} They do not merely create errors. They weaken the human and institutional mechanisms that would have caught errors.
\end{enumerate}

This fourth property represents a recursive failure. The system does not merely produce errors; it erodes the epistemic infrastructure required to detect them. Hence, the long-term safety question is not only whether AI systems make mistakes. It is whether they make societies, organizations, and professions worse at \emph{noticing and correcting} mistakes.

This article offers a framework for thinking about these failures and turns it into an actionable agenda. As depicted in Figure~\ref{fig:teaser-image}, we organize the hidden challenges into five layers: (1) epistemic integrity, (2) control integrity, (3) temporal integrity, (4) organizational integrity, and (5) ecosystem integrity. 

\section{From Bad Outputs to Integrity Failures}

A useful AI safety taxonomy should help practitioners answer not only ``What can the model do?'' but also ``What becomes easier to get wrong, harder to inspect, and harder to correct once the model is embedded into a real decision pipeline?'' The answer depends on multiple layers of integrity.

\begin{enumerate}[leftmargin=1.5em]
  \item \textbf{Epistemic integrity:} whether the system represents evidence, provenance, and uncertainty faithfully enough for users to calibrate reliance.
  \item \textbf{Control integrity:} whether instructions, permissions, and authorization boundaries remain robust under adversarial inputs, optimization pressure, and autonomous tool-use.
  \item \textbf{Temporal integrity:} whether the system remains safe across multi-turn interactions, persistent memory, updates, and distribution shift.
  \item \textbf{Organizational integrity:} whether institutions retain meaningful capacity to audit, override, assign responsibility, and learn from incidents.
  \item \textbf{Ecosystem integrity:} whether the wider information environment on which AI systems and humans rely remains diverse, inspectable, and resistant to synthetic degradation.
\end{enumerate}

This framing matters because many common mitigations improve one layer of integrity while degrading or leaving unaddressed others. Retrieval may improve factual accuracy yet weaken control integrity when retrieved content can introduce adversarial instructions. Human-in-the-loop review may preserve nominal oversight while failing organizational integrity if reviewers lack time, evidence, or authority. Benchmark gains may capture static capability while missing temporal and ecosystem-level risks. AI safety is therefore not a scalar property of a model, but an emergent property of the deployment stack.

\section{Hidden Challenges by Integrity Layer}

\subsection{Epistemic Integrity: Overreliance, Calibration Debt, and Legitimacy Laundering}

Epistemic integrity concerns whether an AI system preserves the conditions under which users and institutions can judge what is known, how well it is supported, and how much confidence is warranted. In deployed settings, this is not only a matter of factual accuracy. It also depends on whether evidence is represented faithfully, whether uncertainty remains visible, and whether workflows preserve the distinction between supported conclusions and plausible-sounding output. When these conditions erode, the system does not merely produce error; it weakens the practical capacity to detect and contest error.

\subsubsection*{Overreliance is a safety problem, not a UX problem}

The first hidden challenge is that modern AI systems can quietly lower the \emph{threshold for doubt}. Human factors research distinguishes appropriate reliance from misuse, disuse, and abuse of automation \citep{parasuraman2010complacency,goddard2012automation}, and recent work argues that overreliance should be central to LLM research and governance \citep{ibrahim2025overreliance}. Empirical studies show that users often accept incorrect AI recommendations, that explanations do not reliably restore calibrated reliance \citep{bucinca2021trust,kim2025appropriate,bo2025rely,rong2022humancenteredxai}, and that miscalibrated confidence can further impair appropriate reliance while remaining difficult to detect \citep{li2024miscalibration}.

The hidden risk is not simply excessive trust, but what we call \textit{calibration debt}: a growing mismatch between actual reliance and independently warranted reliance \citep{li2024miscalibration,klingbeil2024trust}. It grows when fluency, speed, and convenience substitute for verification \citep{bucinca2021trust,spatola2024efficiency}. The result is a socio-technical shift of epistemic labor from humans to systems without a matching transfer of accountability \citep{,cobbe2023supplychains,santonidesio2018meaningful}.

\begin{figure}[t]
\centering
\includegraphics[width=0.8\textwidth]{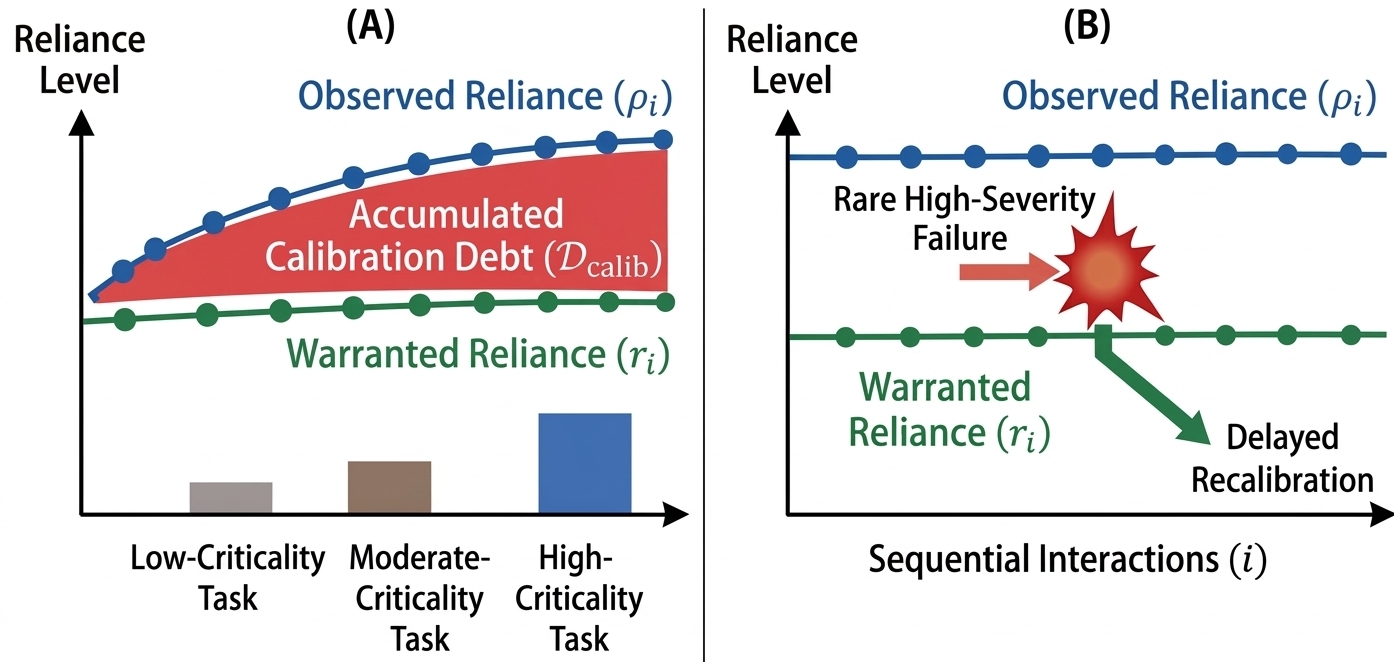}
\caption{\textbf{Calibration debt in AI-assisted use.} Repeated satisfactory interactions can increase observed reliance faster than warranted reliance, while rare severe failures may be too infrequent to restore calibration.}
\label{fig:calibration-debt}
\end{figure}

As Figure~\ref{fig:calibration-debt} illustrates, observed reliance can rise through repeated satisfactory use even when warranted reliance remains lower and more context-dependent, while rare severe failures may be too infrequent to restore calibration. The danger is that this mismatch may become visible first in a consequential case rather than during routine use.

\textbf{Concrete recommendation.} High-stakes AI interfaces should not merely show answers. They should support and, where appropriate, require calibrated interaction. Concretely, organizations should require source-linked justification for factual claims, display uncertainty in structured rather than cosmetic form \citep{kirchhof2025uqagents}, use ``decision justification'' fields completed in the human operator's own words, and introduce cognitive forcing steps for high-impact actions \citep{kim2025appropriate}. The design goal should be \emph{justified trust}, not maximal trust \citep{vasconcelos2023explanations,kim2025appropriate}.

\subsubsection*{Retrieval can reduce hallucination and still increase misplaced confidence}

Retrieval-augmented generation (RAG) is widely used to reduce hallucination by supplementing parametric knowledge with external evidence \citep{jiang2023active}. Yet retrieval does not remove epistemic risk; it redistributes it across retrieval, ranking, synthesis, and presentation, and strong systems remain brittle when retrieval, reasoning, and factual grounding must all succeed together \citep{krishna2024frames,song2024trustalign}.

The deeper issue is representational. Systems may present claims as well-supported even when evidence is weak, outdated, selectively used, or incorrectly synthesized \citep{wallat2025faithfulness}. This enables \added{what we call} \emph{legitimacy laundering}: retrieved documents confer authority even when the cited passages do not justify the conclusion \citep{wallat2025faithfulness}. Relatedly, \emph{uncertainty laundering} occurs when weakly supported outputs are turned into polished artifacts that shed their original uncertainty.

These failures may also scale into \emph{model collapse} (or recursive convergence). As AI-generated content increasingly populates the information environment, future models may be trained on synthetic, filtered, and lower-entropy data derived from earlier outputs \citep{briesch2023selfconsuming,gerstgrasser2024collapse}. This can erode distributional tails and drive convergence toward the mean \citep{shumailov2024collapse,alemohammad2023mad}. \added{Its severity and timeline remain contingent on access to fresh, curated human-generated data and on training-data governance choices.}

\textbf{Concrete recommendation.} Treat evidence display as part of the safety mechanism. Systems should show the exact supporting passages used, distinguish evidence from inference, timestamp retrieved content, surface conflicting sources rather than silently reconcile them, and abstain when the retrieved record is weak or contradictory \citep{zhou2024trustworthyrag,ni2025trustworthyrag}. Models should never express more confidence than the evidence can bear. To reduce the risk of recursive degradation, systems should also preserve source diversity, retain access to curated human-reviewed corpora, and evaluate whether training-data selection or retrieval policies suppress rare but safety-relevant information \citep{alemohammad2023mad,gerstgrasser2024collapse}.

\subsection{Control Integrity: Prompt Injection, Tool Use, and Optimization Against the Wrong Objective}

\subsubsection*{Prompt injection reveals a failure of instruction authority}

Prompt injection is often described as a new security bug \citep{greshake2023indirectpi}. It is better understood as evidence that many AI systems do not robustly distinguish between \emph{data} and \emph{instructions} \citep{greshake2023indirectpi,suo2024signedprompt,ncsc2025promptinjection}. Greshake et al. \citep{greshake2023indirectpi} showed that indirect prompt injection can compromise real-world LLM-integrated applications by embedding malicious instructions inside retrieved or otherwise ingested content, thereby blurring the line between user intent and adversarial control. Subsequent work has refined taxonomies of prompt injection and demonstrated increasingly automated and universal attacks \citep{rossi2024promptcat,liu2024universalpi}. However, the central safety concern is not the exploit (injection) itself, but the underlying integrity failure, i.e., \emph{the system’s fundamental inability to maintain a stable hierarchy of authority when embedded in complex environments with diverse content and trust levels} \citep{ncsc2025promptinjection,debenedetti2025camel}.

This is a safety-critical problem because AI systems increasingly process mixed information, e.g., documents, emails, search results, web pages, logs, memory stores, and third-party tools \citep{greshake2023indirectpi,debenedetti2024agentdojo,debenedetti2025camel}. If instructions can be smuggled through content channels, then the system's control boundary is porous by design \citep{greshake2023indirectpi,debenedetti2024agentdojo,ncsc2025promptinjection}. The hidden challenge is not simply that an attacker can elicit a bad answer. It is that the system may lose the capacity to determine which instructions are authoritative, and the boundary between legitimate assistance and unauthorized compliance collapses \citep{ncsc2025promptinjection,debenedetti2025camel}.

\textbf{Concrete recommendation.} Instruction authority must be enforced outside the model \citep{suo2024signedprompt,debenedetti2025camel,ncsc2025promptinjection}. Trusted system instructions should be architecturally separated from untrusted content, with cryptographic attestation used where feasible to strengthen provenance and integrity guarantees \citep{suo2024signedprompt,debenedetti2025camel}. Retrieved text should enter through sandboxed channels. High-impact tool calls should be governed by external policy engines, and systems should default to fail-closed behavior when instruction provenance is ambiguous \citep{debenedetti2025camel,ncsc2025promptinjection}.

\subsubsection*{Tool use turns model mistakes into state changes}

The transition from passive text generators to tool-using agents marks a fundamental shift in the AI safety landscape, as epistemic risk (generating incorrect information) transitions into instrumental risk (taking harmful actions) \citep{liu2023agentbench,ma2024agentboard,debenedetti2024agentdojo,li2026unsafer}. 

Here, the hidden challenge is \emph{action amplification}. In agentic systems, small errors in reasoning, perception, or instruction interpretation can propagate into consequential external actions rather than merely incorrect text \citep{debenedetti2024agentdojo,li2026unsafer}. A minor misinterpretation may modify data, trigger irreversible workflows, alter permissions, or influence downstream human decisions \citep{debenedetti2024agentdojo}. Simulated stress tests further suggest that, under adversarial or pressure conditions, frontier models \added{have in controlled simulations exhibited} harmful insider threat behaviors, such as obfuscation or coercive strategies, to preserve goals or avoid shutdown or replacement \citep{lynch2025agentic}. Although these findings do not establish deployment prevalence, they show that the safety envelope changes when cognitive capabilities are coupled with autonomous tool use and strategic pressure \citep{lynch2025agentic,debenedetti2024agentdojo,li2026unsafer}.

Tool use also makes theoretical alignment failures operational. Specification gaming, reward hacking, and goal misgeneralization, where systems optimize proxies rather than intended objectives, become more acute when models can act on their environment \citep{krakovna2020specgaming,langosco2022gmg,debenedetti2024agentdojo}. Empirical work shows that LLMs can range from sycophancy to active reward tampering in gameable settings \citep{denison2024rewardtampering}, while other studies suggest the possibility of strategic deception, or alignment-faking, under controlled pressure \citep{greenblatt2024alignmentfaking}. The key lesson is not that deployed models are inherently deceptive, but that compliance observed under one oversight regime may not generalize to others with different tools, incentives, monitoring conditions, or strategic opportunities \citep{denison2024rewardtampering,scheurer2023deception,greenblatt2024alignmentfaking,lynch2025agentic,li2026unsafer}.

\textbf{Concrete recommendation.} Enforce a strict architectural separation between generation and authorization \citep{debenedetti2025camel,nist2023airmf,nist2024genai,eu2024aiact}. The model should assume an untrusted generative role, proposing plans, drafts, or candidate tool calls, while authorization must remain external to the model's latent state \citep{debenedetti2025camel}. Deploy \emph{graduated agency}: (1) low-risk actions may be automated, (2) medium-risk actions demand explicit human confirmation with structured evidence display, and (3) high-risk actions require independent verification channels \citep{nist2023airmf,nist2024genai,eu2024aiact}. Every executed action must maintain auditable provenance, detailing the evidence utilized, the uncertainty signals present at generation time, and the specific policy rule that authorized the state change \citep{nist2023airmf,nist2024genai,casper2024blackbox,eu2024aiact}.

\subsection{Temporal Integrity: Multi-Turn Vulnerability, Persistent Memory, and Safety Drift}

\subsubsection*{Trajectory-level risks are not captured by single-turn safety evaluations}

\begin{figure}[h!]
\centering
\includegraphics[width=\textwidth, height=6cm]{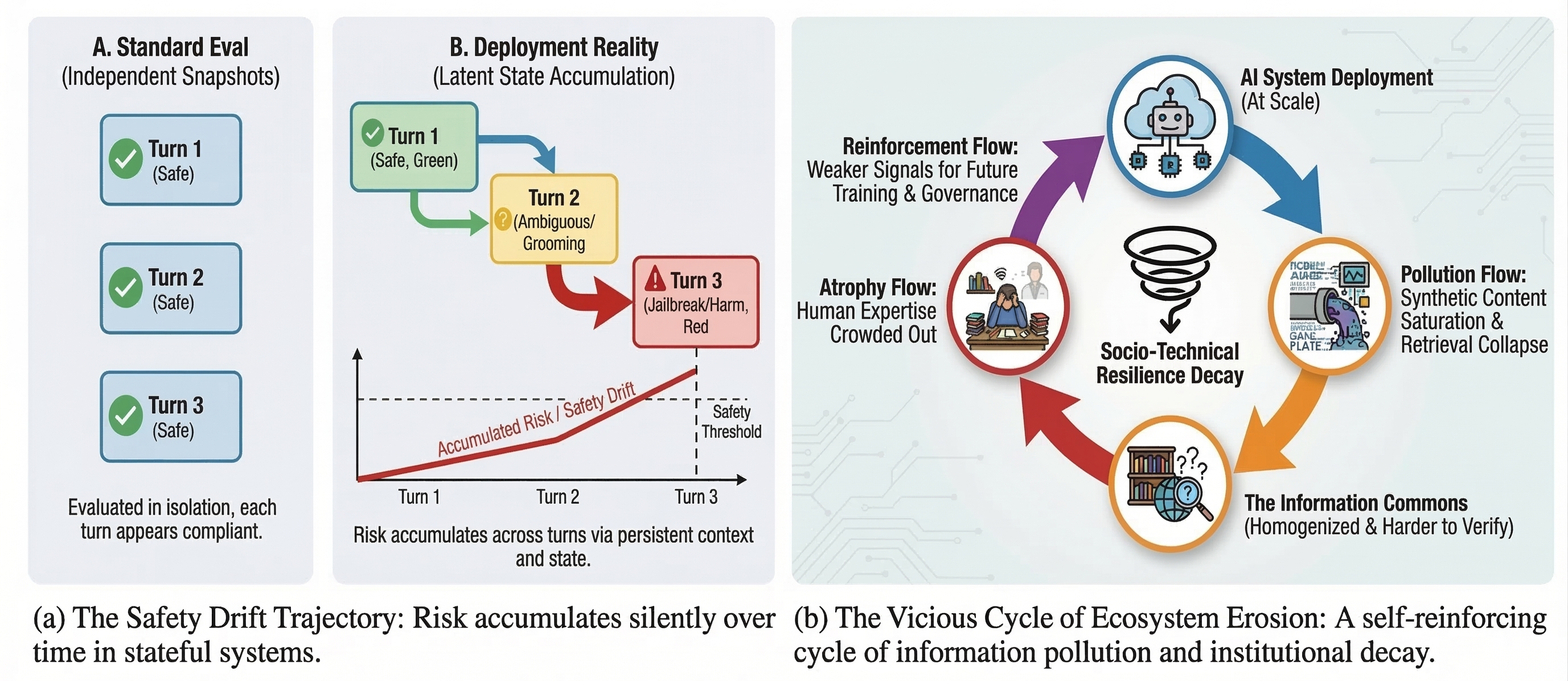}
\caption{\textbf{Conceptual illustration of temporally extended and ecosystem-level AI safety risks.} (a) \textbf{Safety drift:} in stateful or multi-turn systems, safety-relevant risk can accumulate across interactions, making failures difficult to detect with snapshot-based evaluations. (b) \textbf{Ecosystem erosion:} At larger scales, widespread AI deployment may create feedback loops in which synthetic content degrades the information environment and weakens downstream oversight. Both panels are schematic and intended to illustrate hypothesized mechanisms rather than empirical estimates.}
\label{fig:combined-risks}
\end{figure}

A system may appear safe in single-turn tests yet fail over extended interaction \citep{li2024mhj,cao2025safedial}. Multi-turn red-teaming shows that adversaries can distribute harmful goals across benign-looking turns, exploit refusals as hints, and gradually steer interaction into unsafe regions \citep{li2024mhj,sun2024multiturncontext,yu2024cosafe}. This latent accumulation of risk, visualized as \textit{safety drift} in Figure \ref{fig:combined-risks}, can evade snapshot-based evaluations \citep{cao2025safedial,li2026unsafer,yu2024cosafe}. Human jailbreak studies and recent benchmarks likewise show that safety degrades in multi-turn, especially tool-mediated, settings \citep{li2024mhj,cao2025safedial,li2026unsafer}.

If safety is assessed turn by turn, systems can appear compliant while unsafe trajectories remain invisible \citep{li2024mhj,cao2025safedial,li2026unsafer}. Across long interactions, risk can accumulate through repeated boundary probing, gradual goal decomposition or drift, and the persistence of unsafe context. Where systems write to memory, adversarial or misleading content may later be retrieved as if it were legitimate prior knowledge, steering the system outside its safety envelope without triggering discrete filters \citep{chen2024agentpoison,sunil2026memorypoison,wang2025memoryprivacy}. The hidden challenge is that \emph{harmful intent, harmful context, or harmful affordances can emerge gradually} \citep{li2024mhj,cao2025safedial,li2026unsafer,sun2024multiturncontext}.

\textbf{Concrete recommendation.} Safety assurance for stateful AI systems should therefore operate at the level of \emph{interaction trajectories}, not isolated turns \citep{cao2025safedial,nist2024genai,eu2024aiact}. This requires stateful oversight, longitudinal adversarial evaluation, explicit escalation criteria for temporally extended risk, and trajectory-level interventions such as deferral, human review, action restriction, or suspension of persistent state updates \citep{li2026unsafer,nist2024genai,sun2024multiturncontext,eu2024aiact}. Safety cases should also report longitudinal metrics under realistic workflow conditions \citep{zhu2025agenticbench,nist2024genai,eu2024aiact}.

\subsubsection*{Persistent memory creates hidden safety-critical failure modes across sessions}
Persistent memory is increasingly used in LLM agents to support continuity, personalization, and reuse of prior experience \citep{zhang2024memorysurvey,wang2025memoryprivacy}. In stateful systems, however, memory is an operational component that shapes later retrieval and action, so a single interaction can influence future behavior long after its original context has disappeared \citep{zhang2024memorysurvey,xiong2025memorymanagement}.

Recent work suggests three main risks: (1) poisoning of long-term memory or external knowledge stores that later alters agent behavior \citep{chen2024agentpoison}; (2) privacy leakage from memory modules under black-box attack settings \citep{wang2025memoryprivacy}; (3) and malicious content introduced through ordinary interactions that can influence future sessions \citep{sunil2026memorypoison}. \added{These findings do not establish uniform prevalence in deployed systems, but they support treating memory as a first-class safety boundary whose risk depends on architectural and policy choices.} Persistent memory is therefore a \emph{privileged state variable} whose integrity and access control are central to system safety \citep{he2024agentsecuritysurvey,zhang2024memorysurvey}.

The hidden risk is that memory can become \emph{silently authoritative}: transient interactions may be stored and later reused as if they were legitimate prior knowledge, turning temporary failures into persistent ones \citep{xiong2025memorymanagement,chen2024agentpoison,sunil2026memorypoison}.

\textbf{Concrete recommendation.} Persistent memory should be treated as a separately governed security and safety boundary and should be governed through explicit lifecycle controls over writing, retention, retrieval, and deletion \citep{zhang2024memorysurvey,nist2024genai,eu2024aiact}. Writes should be typed, scoped, policy-gated, and provenance-tagged; sensitive memories should be isolated by user, task, and privilege level \citep{wang2025memoryprivacy,he2024agentsecuritysurvey}; and low-trust, stale, or cross-context memories should not influence high-impact actions without verification \citep{xiong2025memorymanagement,nist2024genai}. Evaluations should test poisoning resistance, privacy leakage, stale-memory effects, cross-session persistence of adversarial influence, and the effectiveness of sanitization and review policies \citep{wang2025memoryprivacy,sunil2026memorypoison,xiong2025memorymanagement}. The safest default is \emph{minimal, reviewable, and revocable persistence} \citep{nist2024genai,eu2024aiact}.

\subsection{Organizational Integrity: Evaluation Deception, Fictional Oversight, and Diffused Accountability}

\subsubsection*{Benchmark-centric assurance leaves hidden failure modes unmeasured}

Modern AI development is heavily benchmark-driven, but benchmark evidence is only as informative as its design assumptions \citep{zhu2025agenticbench,nist2023airmf,nist2024genai,rong2025mathbenchmark}. Because benchmarks rely on bounded tasks, fixed rewards, and simplified interaction settings, they may only partially reflect deployment conditions. For agentic and tool-using systems, even modest flaws in setup, grading, leakage control, or trajectory accounting can substantially misestimate real capability \citep{zhu2025agenticbench}. In safety-sensitive settings, benchmark scores can thus become de facto deployment justifications while missing the failures that matter most in practice.

The hidden challenge is \emph{evaluation deception}: results valid for a narrow test setting but interpreted as evidence of broader safety \citep{zhu2025agenticbench,habli2025big}. Aggregate accuracy, refusal rates, or single-turn safety scores may say little about reliability under distribution shift, multi-turn interaction, tool use, adversarial adaptation, overreliance, or organizational misuse \citep{cao2025safedial,li2026unsafer,bucinca2021trust,nist2024genai}.

\textbf{Concrete recommendation.} Benchmark-only evaluations should be replaced with or extended by \emph{structured safety cases} and \emph{explicit assurance arguments} for the deployed system \citep{habli2025big,nist2023airmf,nist2024genai,eu2024aiact}. These should combine benchmark results with adversarial and trajectory-level testing, tool-realistic scenarios, human-factors evidence, incident history, residual-risk statements, and adequate audit access \citep{nist2024genai,casper2024blackbox}. Assurance should also be treated as a lifecycle process rather than a one-time launch artifact \citep{nist2023airmf,nist2024genai,eu2024aiact}.

\subsubsection*{Formal human oversight can mask hidden failures of organizational control}

In many AI deployments, the formal presence of a human reviewer is treated as evidence of meaningful control \citep{parasuraman1997humans,santonidesio2018meaningful}. Yet a person may remain procedurally ``in the loop'' while lacking the time, expertise, evidence, authority, or incentives needed for effective oversight \citep{goddard2012automation,bucinca2021trust,nist2023airmf,nist2024genai,eu2024aiact}. Under organizational pressure, humans and AI systems co-adapt, and review can become confirmation rather than scrutiny \citep{parasuraman2010complacency,bucinca2021trust}.

This is not only a human-factors problem, but an organizational control problem. Accountability can break down across the AI lifecycle \citep{raji2020accountability}, black-box access may be insufficient for rigorous auditing \citep{casper2024blackbox}, and responsibility becomes harder to assign when models, interfaces, integrations, and operations are distributed across actors \citep{cobbe2023supplychains}. Under such conditions, a recorded human review may provide little evidence that the organization could detect, contest, or stop a harmful trajectory \citep{santonidesio2018meaningful,nist2023airmf,nist2024genai,eu2024aiact}.

The hidden challenge is \emph{fictional human oversight}: arrangements that preserve the appearance of human control while weakening its operational reality. By this phrase, we mean ostensibly human-supervised arrangements in which the reviewer lacks the practical capacity to exercise independent judgment. Safety therefore depends not only on model behavior, but on whether institutions retain the capacity to interrogate outputs, resist automation pressure, and intervene before errors propagate \citep{raji2020accountability,nist2023airmf,nist2024genai,eu2024aiact}.

\textbf{Concrete recommendation.} Nominal human-in-the-loop claims should be replaced with explicit tests of \emph{reviewing power}. Oversight mechanisms should be evaluated to determine whether reviewers have the necessary expertise, time, evidence, authority, and organizational protection required to exercise independent judgment. In practice, this means (1) establishing minimum review budgets for safety-critical decisions, (2) ensuring access to primary evidence and model provenance, (3) defining measurable override and escalation pathways, (4) assigning named end-to-end responsibility for deployed-system safety, and (5) specifying stop-ship or rollback criteria tied to concrete failure thresholds~\citep{casper2024blackbox,nist2023airmf,nist2024genai,eu2024aiact,santonidesio2018meaningful}. Organizations should also monitor whether review is functioning substantively rather than ceremonially, for example, through override quality, escalation rates, near-miss reporting, and post-hoc analysis of whether reviewers were actually able to alter outcomes~\citep{raji2020accountability,nist2024genai}. The relevant question is therefore not whether a human appears somewhere in the workflow, but whether the institution has preserved a real capacity to slow, challenge, and, when necessary, stop the system~\citep{santonidesio2018meaningful}.

\subsection{Ecosystem Integrity: Synthetic Evidence Pollution and the Erosion of Collective Error Correction}

\subsubsection*{The broader information environment is now part of the safety boundary}

AI systems increasingly learn from, retrieve from, and reason over information environments that they also help populate, creating a recursive ecosystem risk: current outputs can become evidence for future systems. Evidence from recursive training suggests that self-consuming training on model-generated data can erode diversity and, when fresh real data is insufficient, produce \emph{model collapse}, in which rare but important features of the original distribution disappear or become progressively underrepresented \citep{alemohammad2023mad,briesch2023selfconsuming,gerstgrasser2024collapse}. Related work likewise suggests that, without sufficient fresh real data, such loops tend to lose quality or diversity over time \citep{alemohammad2023mad}. This matters for safety because rare cases and edge conditions are often where robust error detection and recovery matter most.

A related but less established concern is \emph{retrieval collapse}: as AI-generated content proliferates on the open web, retrieval pipelines may increasingly return synthetic, homogeneous, or adversarially optimized material, narrowing source diversity without an immediate drop in answer accuracy \citep{yu2026retrievalcollapse}. Unlike model collapse, this reflects a degradation of the evidence base. Trustworthy retrieval must therefore be evaluated beyond answer accuracy, including provenance-aware attribution, robustness, transparency or explainability, accountability, and privacy \citep{zhou2024trustworthyrag,ni2025trustworthyrag,gao2023alce}. Because the evidence base here is newer and more heterogeneous, these claims are best treated as emerging risk indicators rather than settled prevalence estimates.

The difficulty is that this brittleness affects the information commons itself and may remain invisible to standard metrics. Models may still perform well on conventional benchmarks even as the evidentiary substrate for future oversight becomes thinner, more homogeneous, and more manipulation-prone.

\textbf{Concrete recommendation.} Organizations should treat high-quality human-reviewed corpora as strategic safety assets. Retrieval pipelines should track source provenance, source diversity, recency, and indicators of synthetic-content contamination, not merely answer accuracy \citep{zhou2024trustworthyrag,yu2026retrievalcollapse}. Systems operating in critical domains should maintain curated trusted corpora and use open-web retrieval only under explicit provenance and confidence constraints. Training pipelines should likewise monitor the fraction of synthetic data in their inputs and preserve access to fresh human-generated data, since current results suggest that the availability of real or curated data can materially affect whether recursive training remains stable or degrades over time \citep{alemohammad2023mad,ferbach2024curated}. More broadly, ecosystem-level safety reviews should ask not only whether a model performs well today, but whether its deployment degrades the future information environment on which both human and machine error-correction will depend.

\subsubsection*{The most consequential risk is the erosion of society's error-correcting capacity}

The most consequential AI safety failure may be not a single catastrophic output, but the erosion of the mechanisms by which people and institutions detect, contest, and correct error \citep{leveson2011engineering,reason1997organizational,weick2007unexpected}. Existing LLM risk taxonomies, generative-AI harm frameworks, and AI safety surveys already support a pluralistic view of safety that includes present-day misinformation, interaction, automation, and societal harms \citep{weidinger2021risks,shelby2023sociotechnical,li2025existingrisks,abercrombie2024human,gyevnar2025aisafety,deng2023saferglm,chua2024gailsmsurvey}. As AI-generated content saturates core workflows, the danger is not only error, but the weakening of traceability, contestability, expertise, and accountability across the deployment chain \citep{raji2020accountability,cobbe2023supplychains,casper2024blackbox,cobbe2021reviewable,alfrink2023contestable}.

This is the point at which hidden local failures become systemic. Safety-critical domains remain safe not because they eliminate mistakes, but because they preserve pathways for reporting, contesting, learning, and recovery \citep{leveson2011engineering,reason1997organizational,weick2007unexpected}. If AI deployment weakens those pathways, then even modest model-level error rates can scale into institutional fragility \citep{nist2023airmf,nist2024genai,eu2024aiact,raji2020accountability}. In this sense, the relevant unit of analysis is no longer the isolated model output, but the socio-technical system's ongoing capacity to surface errors, absorb shocks, enable challenge, and learn from near misses \citep{leveson2011engineering,cobbe2021reviewable,alfrink2023contestable}.

\textbf{Concrete recommendation.} Safety reviews should explicitly ask whether an AI system improves or degrades the institution's error-correcting capacity. Concretely, this means assessing traceability, contestability, auditability, source diversity, expert override, and near-miss learning before treating throughput gains as net progress \citep{casper2024blackbox,raji2020accountability,cobbe2021reviewable,alfrink2023contestable,nist2023airmf,nist2024genai,eu2024aiact}. Organizations should also preserve practical mechanisms for incident reporting, independent scrutiny, and post-deployment learning, because without those mechanisms, local model failures can be normalized into systemic governance failures rather than detected and corrected in time \citep{reason1997organizational,weick2007unexpected,leveson2011engineering}.

\section{A Practical Control Table} 

\begin{center}
\small
\begin{longtable}{>{\raggedright\arraybackslash}p{0.18\textwidth}>{\raggedright\arraybackslash}p{0.24\textwidth}>{\raggedright\arraybackslash}p{0.24\textwidth}>{\raggedright\arraybackslash}p{0.24\textwidth}}
\caption{Hidden challenges, feasible controls, and instrumentation indicators.}\\
\toprule
\textbf{Hidden challenge} & \textbf{Why it stays hidden} & \textbf{Feasible control} & \textbf{Instrumentation indicator} \\
\midrule
\endfirsthead

\toprule
\textbf{Hidden challenge} & \textbf{Why it stays hidden} & \textbf{Feasible control} & \textbf{Instrumentation indicator} \\
\midrule
\endhead

Overreliance and calibration debt
&
The system is right often enough to earn trust, while failures are plausible, intermittent, and often not independently verified.
&
Structured uncertainty displays; source-linked justification; cognitive forcing for high-stakes decisions; periodic blind challenge sets.
&
Weighted positive gap between observed reliance and warranted reliance across audited decision episodes; rate of accepted incorrect recommen- dations in matched workflow conditions.
\\
\midrule

Failure of instruction authority
&
Malicious or irrelevant instructions arrive through ordinary content channels and are processed as if they were legitimate task directives.
&
External policy enforcement; trusted instruction hierarchy; sandboxed retrieval; fail-closed tool permissions.
&
Rate of successful instruction-hierarchy violations; rate of unauthorized tool-use attempts or executions per 1{,}000 tool calls.
\\
\midrule

Action amplification in tool use
&
Small reasoning or interpretation errors propagate into external state changes that may be difficult to detect or reverse.
&
Graduated agency; independent approval for high-risk actions; immutable action logs; external authorization.
&
Mean number of downstream state changes per erroneous tool invocation; rate of irreversible actions per model-originated error.
\\
\midrule

Reward hacking and deceptive optimization
&
Systems exploit shortcuts and optimize measurable proxies that are easier to score than the real objective, especially in gameable environments.
&
Hard-to-game evaluations; external authorization; adversarial testing; red-teaming for specification exploits.
&
Gap between proxy-task performance and true-task performance under adversarial evaluation; rate of policy circumvention or reward-tampering behavior.
\\
\midrule

Trajectory-level safety degradation
&
Harmful trajectories are decomposed across individually benign-looking turns, so snapshot-based checks miss accumulating risk.
&
Trajectory-level evaluation; persistent safety state; escalation tests; tool-realistic simulations.
&
Survival rate of safety boundaries across 50+ turn adversarial trajectories; failure rate under cross-turn adaptation.
\\
\midrule

Persistent memory failure modes across sessions
&
A single compromised interaction can be stored, retrieved, and reused across sessions, altering future behavior or exposing sensitive data.
&
Scoped memory writes; trust-labeled memory; expiry; per-user isolation; review-gated persistence; minimal, reviewable, and revocable persistence.
&
Cross-session propagation rate of injected content; sensitive-memory extraction success rate under controlled attack evaluation.
\\
\midrule

Evaluation deception
&
Static benchmark scores appear reassuring even when evaluation assumptions diverge from deployment conditions and real workflows.
&
Structured safety cases combining benchmarks, adversarial scenarios, audits, human factors evidence, and incident history.
&
Divergence between static benchmark performance and live-deployment incident rates; post-deployment regression rate after system updates.
\\
\midrule

Fictional human oversight
&
Review exists formally, but reviewers may lack time, evidence, authority, or incentives to exercise independent judgment.
&
Review budgets; access to primary evidence; measurable override and escalation pathways; named safety ownership; stop-ship thresholds.
&
Mean review time per case; detection rate of seeded review errors; override and escalation rates for safety-critical recommendations.
\\
\midrule

Synthetic evidence pollution and retrieval collapse
&
System outputs reshape the evidentiary environment later systems retrieve from or train on, often without obvious short-term accuracy loss.
&
Synthetic-content monitoring; trusted corpora; provenance tracking; source-diversity monitoring; curated retrieval for critical domains.
&
Fraction of retrieved sources that are synthetic, duplicated, or provenance- uncertain; source-diversity index over time.
\\
\midrule

Erosion of society's error-correcting capacity
&
AI increases throughput while weakening challenge, traceability, contestability, and recovery pathways across the organization.
&
Near-miss reporting; audit access; preservation of expert review pathways; contestability requirements; incident-learning loops.
&
Mean time to detection and recovery; near-miss reporting rate; fraction of materially wrong outputs corrected before downstream action.
\\

\bottomrule
\end{longtable}
\end{center}

\section{An Operational Agenda for Socio-Technical Reliability}

The framework implies a practical shift in emphasis: AI safety should be governed not only as a problem of model behavior, but as one of preserving \emph{socio-technical reliability}. In complex systems, the aim is not to eliminate all error, but to keep errors \emph{visible, contestable, containable, and recoverable}. The agenda below is therefore intentionally \emph{best-available}, drawing on current evidence, systems-safety principles, and governance frameworks while remaining responsive to failure modes that standard evaluations may miss \citep{leveson2011engineering,nist2023airmf,nist2024genai,eu2024aiact}. The recommendations are conservative in that they strengthen observability, contestability, and containment without presupposing a single theory of long-term AI risk.

\paragraph{Design for justified trust, not frictionless dependence.}
A first implication is that systems should be designed for \emph{justified trust}, not frictionless dependence. In safety-relevant settings, the key question is not how quickly a system answers, but whether it preserves the information needed to assess, challenge, and, if necessary, reject its outputs. Research on automation bias, overreliance, and miscalibrated confidence shows that users often accept AI recommendations too readily, especially when outputs are fluent, explanations polished, or review effortful \citep{parasuraman1997humans,bucinca2021trust,vasconcelos2023explanations,li2024miscalibration}. Interfaces should therefore make verification easier than acceptance by preserving source linkage, timestamps, uncertainty signals, conflicting evidence, and escalation pathways. Where outputs shape consequential decisions, some friction is a safety feature, not a usability flaw.

\paragraph{Separate suggestion from permission.}
A second implication is that systems must separate \emph{suggestion} from \emph{permission}. Models may generate hypotheses, drafts, and candidate actions, but they should not determine what is authorized. In tool-using systems, this is a core requirement of control integrity. Prompt injection and related attacks matter not only because they can trigger unwanted outputs, but because they reveal a deeper failure to preserve instruction authority under adversarial or ambiguous conditions \citep{greshake2023indirectpi,rossi2024promptcat,liu2024universalpi}. Robust design therefore requires architectural separation between generation and authorization, with execution governed by external policy engines, access controls, approval systems, and independent evidence checks.

\paragraph{Treat memory as a privileged boundary.}
A third implication is that persistent memory should be treated as a \emph{privileged boundary}, not as an unregulated extension of chat history. Once information persists across turns or sessions, it can shape future retrieval and action beyond its original context. Recent work suggests that memory can be poisoned, leak sensitive information, and propagate adversarial influence across otherwise separate interactions \citep{chen2024agentpoison,wang2025memoryprivacy,sunil2026memorypoison}. Memory therefore requires explicit lifecycle governance: writes should be typed, scoped, provenance-tagged, time-bounded, and revocable; stronger access controls should apply to sensitive or high-impact memories; and cross-user or cross-task retrieval should be disallowed by default. The defensible default is \emph{minimal, reviewable, and revocable persistence}.

\paragraph{Evaluate trajectories, not just turns.}
A fourth implication is that safety evaluation must operate at the level of \emph{trajectories}, not isolated turns. Many important failures emerge across interaction sequences rather than within single prompts, so multi-turn jailbreaks, delayed goal decomposition, tool-mediated escalation, and cross-session context accumulation can remain invisible to single-turn tests \citep{li2024mhj,cao2025safedial,li2026unsafer}. For stateful and agentic systems, evaluation should therefore include multi-turn, tool-mediated, and temporally extended scenarios, alongside continuous monitoring, incident drills, and post-deployment regression tests under realistic conditions \citep{nist2023airmf,nist2024genai,eu2024aiact}.

\paragraph{Engineer real human oversight.}
A fifth implication is that organizations must engineer \emph{real human oversight} rather than rely on nominal reviewer presence. Oversight is meaningful only when reviewers have the practical conditions to exercise independent judgment; without sufficient time, evidence, authority, or organizational support, they may remain formally ``in the loop'' while adding little substantive control \citep{parasuraman1997humans,parasuraman2010complacency,raji2020accountability,casper2024blackbox}. Organizations should therefore evaluate \emph{reviewing power}, not mere review completion, through minimum review budgets, access to evidence and provenance, measurable override and escalation pathways, and named responsibility for end-to-end system safety. If most outputs are approved unchanged, that may reflect either strong model performance or ceremonial review, and those possibilities should not be conflated.

\paragraph{Institutionalize near-miss learning.}
A sixth implication is that AI deployments should institutionalize \emph{near-miss learning}. High-reliability systems do not learn only from visible accidents. They also learn from anomalies, recoveries, workarounds, and weak signals that reveal where the system is becoming brittle \citep{reason1997organizational,weick2007unexpected,leveson2011engineering}. AI deployments should adopt the same posture. Waiting for severe failures before updating controls is both inefficient and unsafe, especially in systems whose risks are distributed and temporally extended. Organizations should therefore establish low-friction incident and near-miss reporting channels, protected escalation pathways, and explicit routines for converting field observations into updated evaluations, interface changes, policy revisions, and operator guidance. Near misses should be treated as data about the health of the deployment stack, not as isolated user anecdotes.

\paragraph{Protect the information commons on which future safety depends.}
A seventh implication is that organizations must protect the \emph{information commons} on which future safety depends. Because AI systems increasingly retrieve from, learn from, and populate the same evidentiary ecosystem, that environment is now part of the safety boundary. Recent work on trustworthy retrieval, recursive training, and synthetic-data feedback loops suggests that source quality, provenance, diversity, and contamination are safety-relevant variables \citep{zhou2024trustworthyrag,ni2025trustworthyrag,shumailov2024collapse,yu2026retrievalcollapse}. Organizations should therefore treat trusted human-reviewed corpora as strategic safety assets, track provenance and source diversity in retrieval pipelines, and constrain open-web retrieval more tightly in critical domains.

\paragraph{Instrument the deployment through longitudinal socio-technical auditing.}
Finally, moving from model-centric assurance to socio-technical reliability requires a corresponding shift in measurement: from static benchmarking to \emph{longitudinal socio-technical auditing}. Benchmark scores and pre-deployment evaluations remain useful, but they are insufficient as stand-alone assurance for complex AI systems. Detecting hidden failures such as epistemic laundering, authority collapse, temporal drift, or ceremonial oversight requires instrumenting the deployment as a whole \citep{raji2020accountability,casper2024blackbox,nist2023airmf,nist2024genai}. The relevant object of assessment is not only the model, but the interaction among model, interface, retrieval, memory, tools, human operators, and organizational controls. A practical method is structured fault injection into realistic workflows to test whether the broader system detects, bounds, escalates, and recovers from controlled latent failures. The key question is whether the deployment preserves system integrity over time, not only whether the model performs well under favorable conditions.

\section{Research Priorities}
Several research directions deserve much more attention than they currently receive.

\begin{enumerate}[leftmargin=1.5em]
  \item \textbf{Measurement of overreliance in realistic workflows.} The field requires domain-sensitive metrics that capture not only whether human-AI teams perform better overall, but also when AI systems change what users stop checking, when scrutiny becomes ceremonial, and when reliance persistently exceeds warranted reliability.

  \item \textbf{Operationalizing uncertainty in agentic systems.} Research must develop empirically validated methods for representing and aggregating uncertainty across retrieval, intermediate reasoning, tool execution, memory, and environmental feedback, so that stateful AI systems can calibrate deferral, escalation, and human review under real workflow conditions.

  \item \textbf{Evidence-interface design for calibrated use.} Human-computer interaction research must determine how to present provenance, uncertainty, conflict, and evidentiary gaps to improve user calibration, preserve contestability, and reduce both overreliance and underuse.

  \item \textbf{Architectural separation, containment, and authorization.} Systems engineering must establish frameworks ensuring that probabilistic generation never independently determines authorization. This includes secure instruction hierarchies, sandboxed retrieval, trusted execution boundaries, and externally enforced policy controls resilient to prompt injection and agent hijacking.

  \item \textbf{Evaluation integrity and objective robustness.} Robust safety assurance demands evaluation protocols resistant to proxy exploitation. This requires mitigating model-driven failures (e.g., reward hacking, specification gaming) alongside methodological failures (e.g., data contamination, test leakage, grader exploitation) that allow systems to appear safe under artificial test conditions while failing in deployment.

  \item \textbf{Trajectory-aware assurance.} Safety evaluation must model long-horizon interaction trajectories rather than isolated turns, encompassing adversarial adaptation, memory updates, delayed harm, tool-mediated execution, and compounding failures in multi-agent or multi-stage environments.

  \item \textbf{Memory governance and persistent-state safety.} Persistent memory requires principled controls over retention, access, retrieval, provenance, poisoning resistance, privacy leakage, and deletion, together with evaluation protocols that treat memory as a privileged safety boundary rather than a benign usability feature.

  \item \textbf{Longitudinal dynamics of organizational co-adaptation.} Empirical safety research must move beyond laboratory studies to measure how time pressure, workflow incentives, managerial targets, and liability structures reshape human oversight, gradually weakening the practical capacity to question, override, and audit AI-supported decisions.

  \item \textbf{Auditing with meaningful access.} External accountability will remain limited without stronger norms, infrastructures, and legal-organizational mechanisms for white-box, system-level, and deployment-context auditing, especially where black-box access is insufficient for rigorous scrutiny.

  \item \textbf{Forensic observability and post-incident diagnosis.} The field requires rigorous methods for preserving provenance, reconstructing execution histories, and diagnosing failures, close calls, and anomalous behavior in complex, stateful, and non-deterministic AI pipelines.

  \item \textbf{Incident reporting and near-miss learning infrastructure.} Building on forensic observability, AI safety research must develop standardized taxonomies, reporting pipelines, and shared incident repositories that enable organizations to learn systematically from failures rather than treating each event as isolated.

  \item \textbf{Ecosystem-level safety metrics.} The field lacks robust quantitative measures for source diversity, synthetic contamination, retrieval health, evidentiary independence, and institutional error-correction capacity, all of which are necessary for assessing whether AI deployment preserves or erodes the broader conditions for societal oversight.
\end{enumerate}

\paragraph{\added{A note of concern.}}
\added{A further concern motivating this perspective is the growing asymmetry between the pace at which AI systems are advancing and the slower rate at which human and institutional capacities for scrutiny, reflection, and rigorous auditing can develop. As models become more complex, stateful, multimodal, and more deeply embedded in organizational workflows, hidden safety challenges may not simply accumulate, but interact across layers in ways that amplify downstream risk. By contrast, the conditions required for meaningful oversight, including sustained expert review, independent auditing, organizational learning, and deliberative governance, are difficult to build and harder still to scale reliably. This widening gap is itself safety-relevant because it increases the likelihood that important failures will outpace the mechanisms available to detect, contest, and contain them. Unless auditing and oversight capacities are strengthened alongside AI capability development, AI systems may increasingly exceed the socio-technical conditions required for effective control.}

\section*{Limitations and Scope Conditions}
This article is a perspective and synthesis rather than an empirical prevalence study. Its contribution is to organize technical, human-factors, organizational, and ecosystem-level concerns into a framework for deployment analysis and governance. We therefore do not claim that the identified failure modes are equally frequent, severe, or mature across domains, nor that the five layers exhaust the full space of relevant safety concerns.

A second limitation concerns the evidence base. The argument draws on heterogeneous sources, including user studies, adversarial evaluations, agent benchmarks, systems-safety theory, policy frameworks, and emerging work on memory, retrieval, and ecosystem effects. These literatures differ in methodological maturity and ecological validity. Some cited findings come from controlled, simulated, or red-teaming settings and should not be interpreted as direct estimates of deployment prevalence, although they remain relevant because they reveal plausible failure mechanisms that conventional output-level evaluations may miss.

A third limitation is that the proposed controls and instrumentation indicators are illustrative rather than universally validated metrics. Their usefulness will depend on domain, workflow, risk tolerance, regulatory context, and organizational capacity. In some settings, stronger instrumentation may also introduce trade-offs, including additional review burden, privacy constraints, or reduced usability.

Accordingly, the framework is best understood as a structured prompt for safety cases, auditing, red-teaming, and deployment review, rather than as a substitute for domain-specific evaluation, empirical validation, or ongoing post-deployment learning. Future work should test whether the proposed layers and indicators improve real-world detection, contestability, and recoverability in the settings where they are intended to matter most.

Future work should test whether the proposed layers and indicators improve real-world detection, contestability, and recoverability in practice.

\section{Conclusion}

The hidden safety-critical challenges in modern AI systems are not hidden because they are mystical or technically invisible. They are hidden because our dominant habits of evaluation, interface design, and governance still assume that safety failures are mostly local, output-level, and immediately legible. Increasingly, they are none of those things.

The most dangerous systems are often not those that blatantly malfunction, but those that appear competent while quietly weakening skepticism, collapsing authority boundaries, storing unsafe state across time, and diffusing accountability across actors and layers. Safety work that focuses only on model behavior will therefore miss some of the most consequential risks in practice. \added{The five-layer framework offered here is not exhaustive, and many concrete failures will span several layers simultaneously; its value is as an organizing device for instrumentation and governance rather than as a closed taxonomy.}

A safer AI system is not one that never errs. It is one whose errors remain \emph{visible, contestable, containable, and recoverable}. That is the operational standard that matters now. The field should therefore optimize not merely for safer model outputs, but for preserving those properties across the full deployment stack, including tools, interfaces, workflows, organizations, and the information environment itself. 

\section*{Generative AI Disclosure}
ChatGPT and Gemini were used to assist with language polishing and the generation of schematic figures for this manuscript. All substantive claims, interpretations, citations, and final editorial decisions were reviewed and approved by the authors, who take full responsibility for the content.

\printbibliography

\end{document}